\documentclass[aps,twocolumn,groupedaddress,showpacs]{revtex4}
\usepackage{graphicx}
\usepackage{float}
\usepackage{dcolumn}
\usepackage{amsmath}
\usepackage{amssymb}
\input{epsf}

\newcommand{\qedsymb}{\hfill{\rule{2mm}{2mm}}}

\newcommand{\ket}[1]{\left| {#1} \right\rangle}
\newcommand{\bra}[1]{\left\langle {#1}\right |}

\newcommand{\spinhalf}{spin-$1 \over 2$}

\def\p{\ket{\psi}}
\def\Tr{{\rm Tr}}

\def\be{\begin{equation}}
\def\ee{\end{equation}}
\def\bea{\begin{eqnarray}}
\def\eea{\end{eqnarray}}

\def\>{\rangle}
\def\<{\langle}

\begin{document}
\title{
\[ \vspace{-2cm} \]
\noindent\hfill\hbox to 1.5in{\rm  } \vskip 1pt \noindent\hfill\hbox
to 1.5in{\rm SLAC-PUB-13096 \hfill  } \vskip 1pt
\noindent\hfill\hbox to 1.5in{\rm \today \hfill}\vskip 10pt
Bootstrap Approximations in Contractor Renormalization
\footnote{This work was supported by the U.~S.~DOE, Contract
No.~DE-AC02-76SF00515.}}
\author{M. Stewart Siu\footnote{msiu@stanford.edu} and Marvin Weinstein\footnote{niv@slac.stanford.edu}}
\address{Stanford Linear Accelerator Center, Stanford University,
  Stanford, California 94309}

\begin{abstract}

We propose a {\it bootstrap\/} method for approximating the long-range terms in the
Contractor Renormalization (CORE) method. The idea is tested on the 2-D Heisenberg antiferromagnet
and the frustrated $J_2-J_1$ model.  We obtain renormalization group flows that directly reveal the
 Neel phase of the unfrustrated HAF and the existence of a phase transition in the $J_2-J_1$ model
 for weak frustration. However, we find that this bootstrap method is dependent on blocking and truncation schemes.  For this reason, we discuss these dependencies and unresolved issues that researchers who use this approach must consider.

\end{abstract}

\maketitle


\section{Introduction}

Contractor Renormalization (CORE) \cite{COREpaper} is a numerical method for finding the low energy spectrum
 and generating the renormalization group flow of operators in quantum many-body models. Although it has been
 applied successfully to many systems (see references in Ref.\cite{haf1}, and for recent works, Refs.\cite{capponi2,recentCORE}), it relies on a cluster expansion
 whose higher order terms are, in most cases, difficult to compute. However, as we noted in Ref.\cite{haf1}, in
 order to have confidence that a CORE computation is reliable, the convergence of the cluster expansion has to be
  checked carefully.  This is particularly true for higher-dimensional systems.  Unfortunately, for even the
  simplest 2-D systems, the computation of operators beyond nearest-neighbor blocks is a formidable task since, if
   we suppose the number of states per block is $m$, and the cluster
expansion up to the $k$-th term is defined to include all connected
operators in a $k\times k$ configuration of blocks, then the brute
force calculation of the $k$-th term would require exact
diagonalization of an $m^{k^2}\times m^{k^2}$ matrix. Clearly this
is not feasible for large values of $k$.  This paper reports our
attempts to devise schemes for approximating these longer-range
terms in the cluster expansion.

Getting a reasonably reliable handle on longer-range terms is important for two reasons.  First, we would like to obtain the best accuracy for the ground-state and excited state energies, as well as the best values for critical exponents and other order parameters.  Second, and possibly more important, higher order terms will generate long-range operators that can introduce frustration, and therefore possibly produce qualitative changes in the renormalization group flows.  In this paper we propose a {\it CORE bootstrap\/} scheme, wherein we use a variation of CORE itself to compute approximations to these long-range terms.
The key idea behind this approach is that, since our focus is on the computation of new connected operators for a small number of blocks, we use a block adapted version of CORE to approximate the cluster Hamiltonian.  By block adapted we mean that we use a truncation and renormalization procedure specifically tailored to the finite size and geometry of the particular configuration.

The outline of this paper is as follows. In Section
\ref{sec:moreblock}, we first test a variety of bootstrap schemes
for the case of an unfrustrated antiferromagnet(HAF), since the
ground state energy density of this model is known
exactly\cite{exact}.  We show that the increase in accuracy we
obtain for the ground state energy density can show significant
dependence upon the blocking scheme.  Fortunately, all schemes
consistently reproduce a renormalization flow that illuminates the
N\'{e}el nature of the HAF ground state. We should remark that these
schemes can be used to calculate observables such as the staggered
magnetization in the manner shown in Ref.\cite{haf1}. Adding
observables to the Hamiltonian, however, typically breaks some
symmetries and significantly complicates the computation. Therefore
we only focus on the energies and the couplings in this exploratory
work of the bootstrap idea.

In Section \ref{sec:j1j2}, we take the best blocking scheme from the
previous section and apply it to a specific frustrated
antiferromagnet, the $J_1-J_2$ model. This scheme predicts a
renormalization flow as a function of frustration, in which a phase
transition appears around $0.37<J_2/J_1<0.38$. This is in rough
agreement with literature, and it is the first time such a
renormalization picture is obtained for this well-studied model.
Section \ref{sec:other} reports on the results of some variations of
this bootstrap procedure and the puzzles they pose. Finally, we
conclude Section \ref{sec:disc} with a discussion of what we believe
the implications of these results to be.

\section{Blocking Schemes for the Bootstrap Step}
\label{sec:moreblock}

As in Ref.\cite{haf1}, we begin our study of the CORE bootstrap by
studying the 2-D spin-1/2 Heisenberg Antiferromagnet: \bea
H&=&\sum_{<i,j>} \vec{S}_i\cdot \vec{S}_j \nonumber \\
&=&\sum_{<i,j>} S^x_i \cdot S^x_j + S^y_i \cdot S^y_j + S^z_i \cdot
S^z_j ,
\eea
As before, we assume nine-site square blocks for the basic CORE renormalization group computation and truncate to the two lowest lying eigenstates of the corresponding nine-site Hamiltonian.  The simplicity of the lowest range CORE computation for this system makes it a good test bed for a computation of the four-block plaquette operator.  Note that the CORE prescription for computing this operator exactly would be require computing those lowest lying eigenstates of the $2^{36}\times 2^{36}$ cluster Hamiltonian which have a non-vanishing overlap with the $16$ tensor product states constructed in the truncation step. In general, this means computing more than the $16$ lowest lying eigenstates of the $36$-site problem. This is a formidable problem and it cannot be done on a PC.

\begin{figure}
  \begin{center}
  \includegraphics[width=2in]{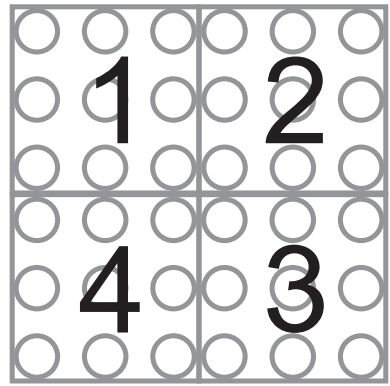}
  \end{center}
  \caption{The labeling of the four nine-site blocks.}
\label{notation}
\end{figure}

A simplification which follows from the fact that our renormalization procedure preserves the rotational invariance of the original theory, is that the renormalized Hamiltonian must be rotationally invariant.  This means that, since we truncate to the lowest spin-${1\over 2}$ state per block, the generic two-block and the generic plaquette operators that appear in the Hamiltonian will have the form:
\bea
h_2 &=& c_{uu} + 4 c_{xx} \vec{S_1} \cdot \vec{S_2} \\
h_4&=&c_{uuuu} \mathbf{1} + 4 c_{xxuu} (\vec{S_1} \cdot \vec{S_2} +
\vec{S_2} \cdot \vec{S_3} \nonumber\\
&&+ \vec{S_3} \cdot \vec{S_4} + \vec{S_4} \cdot \vec{S_1})+4
c_{xuxu} (\vec{S_1} \cdot \vec{S_3} + \vec{S_2} \cdot \vec{S_4})
\nonumber\\
&& + 16 c_{xxzz} (\vec{S_1} \cdot \vec{S_2} ~~\vec{S_3}
\cdot \vec{S_4} + \vec{S_1} \cdot \vec{S_4} ~~\vec{S_2} \cdot
\vec{S_3}) \nonumber \\
&& + 16 c_{xzxz} (\vec{S_1} \cdot \vec{S_3} ~~\vec{S_2} \cdot
\vec{S_4}) \label{h4} \eea where the $36$ sites in the four
nine-site blocks are labeled as shown in Fig.\ref{notation}.  The
constants appearing in Eq. \ref{h4} are put in to ensure that the
coefficients $c_{xzxz}$ are in front of operators whose trace norm
is unity. The coefficients $c_{xuxu}, c_{xzxz}$  indicate the
strength of diagonal interactions, while $c_{xxuu}, c_{xxzz}$ are
for horizontal/vertical interactions. Note that both the $c_{xx}$
and $c_{xxuu}$ terms are $\vec{S} \cdot \vec{S}$ on two neighboring
sites, so the total horizontal or vertical coupling should be
$(c_{xx}+2 c_{xxuu})$. Since the overall scale is irrelevant when we
study the flow of couplings, in the plots to follow we will often
express the couplings in Eq. \ref{h4} in units of $(c_{xx}+2
c_{xxuu})$.
Given this notation, we now turn to a discussion of several possible bootstrap schemes: the Whole Block Buffering scheme, the Three-site Strip Blocking scheme, the Contour Line Blocking scheme and the "8+1" Blocking scheme.

\subsection{A First Attempt: Whole Block Buffering}
\label{sec:whole}

The key to the {\it CORE bootstrap\/} is the introduction of an intermediate, or buffering, step between each of the original CORE renormalization group steps.  In this intermediate step we retain, along with the original \spinhalf\ doublet retained in the first step, additional spin multiplets so that the final number of retained states per block is $2^9 > k > 2$.  This means that for a $4$-block cluster there will be a total of $k^4$ retained states.

Next, we use the CORE contraction procedure to construct a $k^4 \times k^4$ Hamiltonian which approximates the full $2^{36} \times 2^{36}$ in the following sense: one its low lying states should have nearly the same eigenvalue spectrum as in the full $36$-site problem; two, if one restricts to the lowest two states of the single block Hamiltonian, then the renormalization group step should produce a reasonable approximation to the connected operators we would obtain by doing the exact $36$-site computation.  To be precise, this approximating $k^4 \times k^4$ Hamiltonian is constructed as follows:  first, we compute the $k \times k$ range-$1$ (single block) terms and $k^2 \times k^2$ connected range-$2$ (two adjacent block) terms; second, we add these terms up to construct the approximate Hamiltonian.  The hope is that since, if we kept all $2^9$ states per block, there would be no plaquette operator at all, keeping the range-$2$ approximation for $k > 2$ states will, for big enough $k$, be a reasonable thing to do.  The only question is how big does $k$ have to be for this to be true.

An important feature of this procedure is that we keep the original \spinhalf\  multiplet as one of our $k$ retained states.  This means that, if we carry out a nearest neighbor renormalization group step for this approximate Hamiltonian, keeping only the lowest \spinhalf\ multiplet for a single site, then we will automatically reproduce the result obtained in the full eighteen-site, two-block calculation.
Obviously, this procedure can be generalized in ways that fail to guarantee that the connected range-$1$ and connected range-$2$ terms computed for the effective problem will be the same as those obtained from the original $18$-site calculations.  In that case, however, the question of what is the proper definition of the connected plaquette operator raises its ugly head.  To avoid this complication, in what follows, we will only discuss buffering stages for which this cannot happen.

\begin{figure}
  \begin{center}
  \includegraphics[width=3.3in]{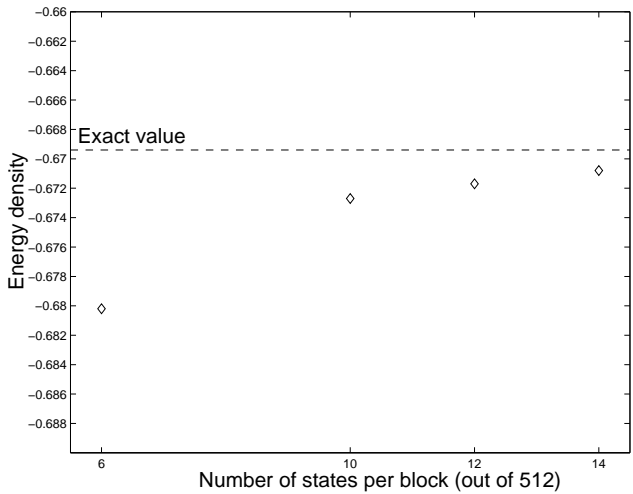}
  \end{center}
  \caption{Approximate energy density as a function of $k$, the number of states per block in the buffering step.
   The four $k$ values are chosen to preserve whole spin multiplets in
the first iteration. For subsequent iterations, we keep instead all
the whole multiplets that can be fit within the $k$ lowest states.}
\label{long}
\end{figure}

The purpose of the buffer step is to reduce numerical effort, so we
limit ourselves to calculating range-$2$ configurations - those
connecting horizontal, vertical and diagonal nine-site blocks. A
point worth noting is that after the first CORE iteration our
renormalized Hamiltonian will contain a connected plaquette term
which couples every four site square. Limiting ourselves to
range-$2$ means the $c_{xxzz}, c_{xzxz}$ terms at the very center of
the 36-site cannot be included. While these operators have very
small norms, this approach also violates the diameter expansion rule
proposed in Ref.\cite{haf1} by not including all diameter-$\sqrt{2}$
operators. This is potentially a defect of the whole block buffering
scheme. Nevertheless, for the purpose of testing and comparison, we
follow this strategy and see what happens for several values of $k$.
The resulting energy densities are shown in Fig.\ref{long}. The
general question of how best to perform whole block buffering
without discarding any terms is an issue that we may return to at a
later date.

Fig.\ref{long} shows that as we add more states, we get closer and
closer to the exact energy density. This is nontrivial and appears
to indicate that we are on the right track, but notice that for
$k=6$, the energy density at $\sim-0.68$ is much worse than the
$-0.666$ obtained without bootstrap \cite{haf1}. This may be the
result of the defect mentioned above.

\begin{figure}
  \begin{center}
  \includegraphics[width=2in]{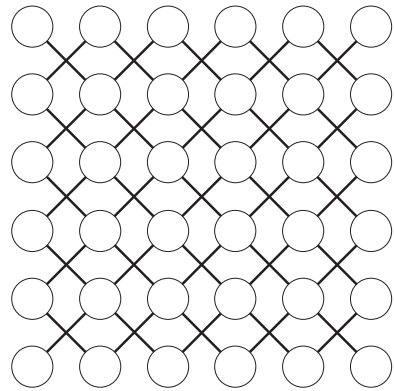}
  \end{center}
  \caption{Two decoupled inter-penetrating lattice. The fact that the HAF runs toward this configuration
  illustrates the N\'{e}el character of the ground state.}
  \label{interp}
\end{figure}

Apart from the energy density, this calculation (as well as all
other blocking schemes to be discussed) reveals an important feature
of the HAF.  With each successive iteration, the sum of
$c_{xx}+2c_{xuxu}$ goes to zero and $c_{xuxu}$ becomes more and more
negative. At the third iteration of $k=14$, for example, we have the
following coefficients: ${c_{xuxu}}/(c_{xx}+2c_{xuxu})\approx-74,~
{c_{xxzz}}/{(c_{xx}+2c_{xuxu})}\approx0.006,~
{c_{xzxz}}/{(c_{xx}+2c_{xuxu})}\approx-4.$ This means that under the
renormalization group flow our lattice Hamiltonian flows to that of
two weakly-coupled interpenetrating ferromagnets (Fig.\ref{interp}).
This picture appears to provide a very attractive qualitative
explanation as to why, at long range, the HAF ground state is
N\'{e}el-like.

\subsection{Three-site Strips}
\label{sec:strip}

The previous blocking scheme leaves the nine-site blocks intact, but nothing prevents us from breaking them up in the buffering step. Appropriately dividing the block into smaller pieces allows us to include more interactions at a smaller computational cost.  More importantly, this flexibility allows us to retain four-body interactions we previously discarded, and so we will do not have to violate the idea of ordering the cluster expansion by diameter/proximity.

\begin{figure}
  \begin{center}
  \includegraphics[width=3.4in]{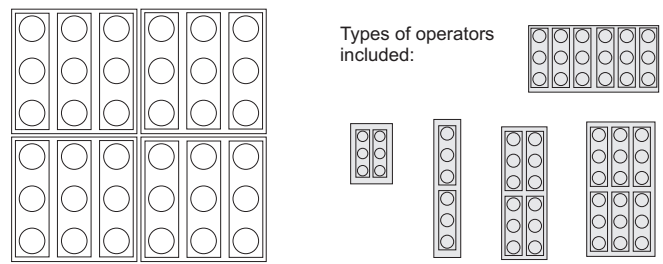}
  \end{center}
  \caption{Three-site strip blocking. Shown on the right are operators for the buffering step that can be embedded in the
  36 sites that involve no more than the diagonalization of two original nine-site blocks.}
  \label{breakblock}
\end{figure}

Fig.\ref{breakblock} shows a way of breaking the nine-site block
into three rectangular strips, along with a list of operators
calculated in the buffering step.  Embedding these terms in the
$36$-site problem, we see that there are two six-strip terms that
connect all four nine-site blocks.  For this strip blocking we keep
two states (a spin-$\frac{1}{2}$) per strip, which means that a
six-strip term requires contracting the lowest of $2^{18}$ states to
$2^6$ states. Computationally this is not much more demanding than
the $2^{18}$-to-$2^2$ contraction required for the original
range-$2$ calculation without bootstrap.  Thus, in this way we can
take into account the four-block interaction more satisfactorily
than in the whole block buffering scheme.

The obvious problem with this approach is the loss of rotational
symmetry.  Now, the up-down couplings will, in general, be different
from the left-right couplings.  This difference can and will grow as
we run the renormalization group flow.  This, of course, limits the
number of RG steps which can be carried out and so, if the limiting
behavior of the renormalized Hamiltonian hasn't clearly emerged
before these asymmetries grow too large one loses the ability to
understand the long distance behavior of the theory.  We can
mitigate this problem somewhat by alternating the blocking direction
at every iteration, but this only improves the energy from $-0.6678$
to $-0.6679$.  Compared to the exact value at $-0.6694$.  Both
numbers are $\sim 0.2\%$ accurate, as was the $k=14$ result in Fig.
\ref{long}.

\begin{figure}
  \begin{center}
  \includegraphics[width=3.4in]{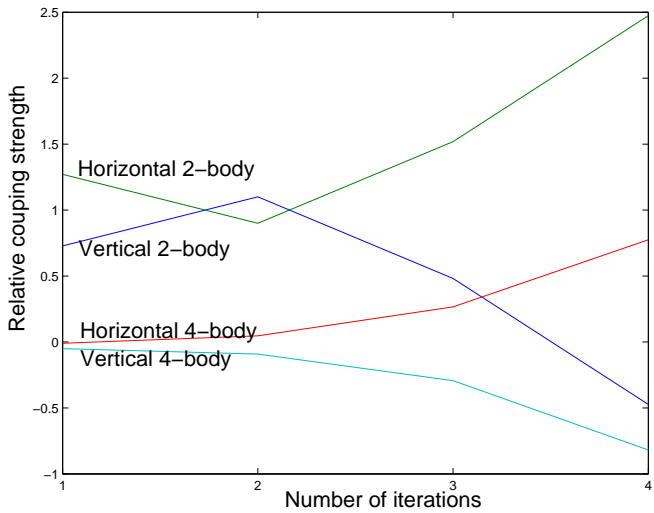}
  \end{center}
  \caption{(Color Online) Running of horizontal/vertical couplings in four CORE iterations using alternating strip blocking. The two-body horizonal
  coupling is defined to be
  $2(c^h_{xx}+2c^h_{xxuu})/(c^h_{xx}+2c^h_{xxuu}+c^v_{xx}+2c^v_{xxuu})$
  and the four-body horizontal coupling is $2c^h_{xxzz}/(c^h_{xx}+2c^h_{xxuu}+c^v_{xx}+2c^v_{xxuu})$. The vertical couplings
  are defined similarly.}
  \label{asymm}
\end{figure}

\begin{figure}
  \begin{center}
  \includegraphics[width=3.4in]{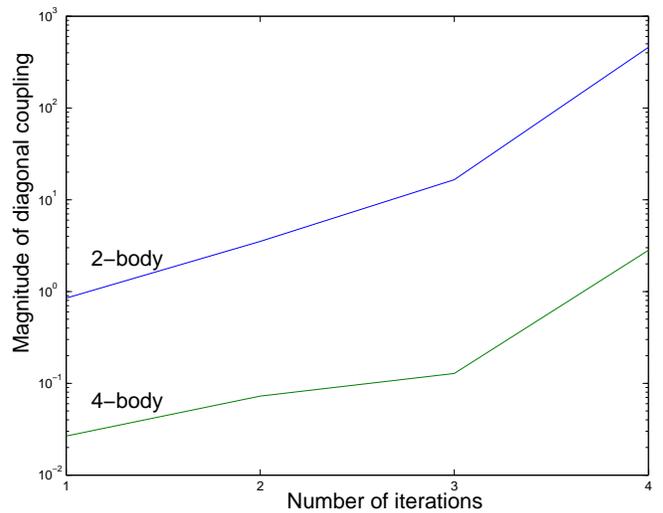}
  \end{center}
  \caption{(Color Online) Running of diagonal couplings in four CORE iterations using strip blocking with averaging of coefficients.
  The two-body diagonal term is $2c^h_{xuxu}/(c^h_{xx}+2c^h_{xxuu}+c^v_{xx}+2c^v_{xxuu})$ and the four-body diagonal
  term is $2c^h_{xzxz}/(c^h_{xx}+2c^h_{xxuu}+c^v_{xx}+2c^v_{xxuu})$. This shows that the HAF runs to the two-lattice limit
  in Fig.\ref{interp}}
  \label{stripavg}
\end{figure}

Fig.\ref{asymm} shows how the asymmetry between horizontal and
vertical couplings grows with renormalization under the alternating
strip-blocking scheme.  Just as in the whole-block case, both
horizontal and vertical couplings are eventually dominated by the
diagonal couplings. To see this more clearly, we artificially
restore the rotation symmetry by averaging the horizontal and
vertical interaction, i.e. set $c^h_{xx} \leftarrow
(c^h_{xx}+c^v_{xx})/2$ etc. where the superscripts indicate
horizontal and vertical coefficients. The averaging scheme hardly
changes the energy density, which now becomes $-0.6677$, but here we
get a clean picture (Fig.\ref{stripavg}) of the growth of diagonal
couplings.  The flow to two weakly-coupled interpenetrating
ferromagnets is once again made manifest.

\subsection{Contour Lines}
\label{sec:contour}

\begin{figure}
  \begin{center}
  \includegraphics[width=3in]{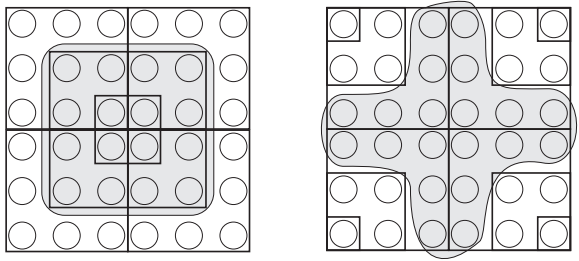}
  \end{center}
  \caption{Two different types of rotationally symmetric contour line blocking. We will refer to the left configuration as a racetrack contour,
  and the right as a cross contour. For the racetrack contour, we have calculated two operators that couples all four nine-site
  blocks; the $4\times 4$ square in the middle (shaded above), and the 20-site outer ring. For the cross contour, the 20-site cross in the middle
  is the term connecting four blocks we can handle.}
  \label{racetrackblock}
\end{figure}

Since three-site strips artificially break the rotational symmetry,
one may wonder if we can rearrange the strips to avoid this. Fig.
\ref{racetrackblock} exhibits two arrangements that achieve this by
breaking up the nine sites into $1+3+5$ sites, forming a contour
line structure.  Notice that this is no longer a scheme which could
be used to compute an RG flow for the full lattice, because no
matter how we orient the contour lines, no single arrangement covers
the entire lattice (four arrangements would be needed).
Nevertheless, in the buffering step there is no barrier to using
either of these arrangements to construct the approximation to the
$36$-Hamiltonian.

\begin{table}
\caption{Performance of strip and contour-line blocking as measured
by energy density. Two states per strip is assumed except for G)}
\begin{tabular}{|l|c|}
  \hline
  Bootstrap Scheme Used  & Approx. Energy Density \\
  & and Percentage Error \\
  & (Exact = -0.66943) \\ \hline
  A) Three-site strip & -0.6678 (+0.24\%)\\ \hline
  B) A) with alternating & \\
  ~~~~orientation & -0.6679 (+0.22\%)\\ \hline
  C) A) with horizontal/vertical & \\
  ~~~~coefficients averaged & -0.6677 (+0.25\%)\\\hline
  D) Racetrack contour without & \\
  ~~~~extra long-range terms & -0.6729 (-0.52\%)\\ \hline
  E) D) with $4\times 4$ square in the& \\
  ~~~~middle included & -0.6658 (+0.54\%\\ \hline
  F) E) with 20-site outer ring &
  -0.6646 (+0.72\%)\\ \hline
  G) Like D) but with four & \\
  ~~~~states instead of two kept in & \\
  ~~~~each five-site strip & -0.6669 (+0.37\%) \\ \hline
  H) Cross contour with 20-site & \\
  ~~~~cross in the middle & -0.6718 (-0.36\%)\\ \hline
  I) Coefficients from F) & \\
  ~~~and H) averaged at the end & \\
  ~~~of each iteration & -0.6668 (+0.39\%)\\ \hline
\end{tabular}
\label{etable}
\end{table}

Table \ref{etable} lists the results of a number of different
contour line blockings.  Blocking scheme D) makes use of the
racetrack contour in Fig.\ref{racetrackblock} and keeps a
spin-$\frac{1}{2}$ multiplet in each strip (made of one, three or
five sites), which is always possible with an odd number of sites.
This scheme does not include any extra long-range terms apart from
the term connecting two adjacent nine-site blocks (made of six
strips), so it is similar to the whole block buffering scheme.
However, instead of including the interaction between two diagonal
nine-site blocks, we only include the four-site interaction in the
middle, and this does not have to be calculated because the four
sites in the middle have been kept free.

Blocking scheme E) takes this further by adding the operator corresponding to the 16 sites in the middle. To make sure we are only adding connected operators, we have to calculate embedded sub-blocks, such as the $2\times 4$ rectangle, etc. Blocking scheme F) includes yet another term corresponding to the 20-site outer ring. Note that there are still many long-range operators that can be embedded in the 36 sites under racetrack contour that we have not included. For instance, we have not calculated the 20-site "T"-shaped term corresponding to two nine-site blocks plus two other free-sites in the middle. However, it would seem that E) and F) already encodes most of the interaction coupling the four nine-site blocks, and this we know is important because the diagonal coupling grows with renormalization.

As apparent from the energy density, our hope was not borne out. The
energy densities obtained from E) and F) are even further from the
exact value than the result without bootstrap \cite{haf1}. To see
whether including more states would help, we modify D) by increasing
the number of states in five-site strips from two to four (i.e. two
spin-$\frac{1}{2}$ multiplets). This is blocking scheme G), and
while it improves the accuracy, it is only about as good as the
non-bootstrap calculation. Switching from the racetrack contour to
the cross contour and including the 20-site for the strips in the
middle (shaded on the right of Fig.\ref{racetrackblock}) gets us
about the same level of accuracy. Finally, artificially averaging
the coefficients from the racetrack contours with those from the
cross contours does not help either.

This experiment appears to teach us two things. First, we see that
the numbers roughly bracket the exact value, so just as in
Fig.\ref{long}, this indicates that a good approximation of the
four-block plaquette should yield a very good energy density.  On
the other hand we learn that a complex blocking scheme is prone to
over-emphasizing certain interactions, and that adding longer-range
operators in such situation does not necessarily lead to better
accuracy. In absence of something resembling the diameter expansion
rule \cite{haf1}(which only makes sense in large, uniform lattices),
it is not clear what operators we should calculate to get
improvements.

\subsection{The "8+1" Blocking Scheme}
\label{sec:8+1}
\begin{figure}
  \begin{center}
  \includegraphics[width=2in]{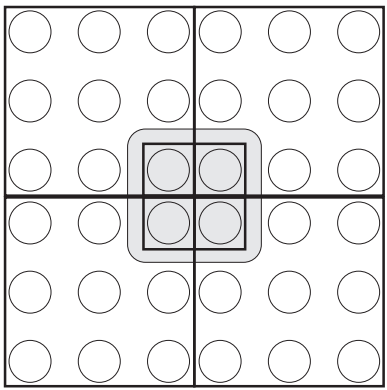}
  \end{center}
  \caption{The "8+1" blocking scheme. One site out of each nine-site block is left free, enabling the interaction
  on the four sites in the middle to be included naturally.}
  \label{fatL}
\end{figure}

With the lessons of our previous attempts in mind, we now present a
blocking scheme that produces much better accuracy.  We saw that the
whole blocks in Section \ref{sec:whole} gave reasonably good results
and did not assume arbitrary structures.  The problem there was that
the four-site operator in the middle, which would be present after
the first iteration or in a frustrated model, cannot be handled in a
simple, natural manner.  The minimal way to deal with this is to
break the nine sites into 8+1 sites, leaving the corner site towards
the middle of the four blocks free. This is shown in Fig.\ref{fatL}.

To mimic the $14$-state whole block calculation, we keep all the multiplets that can be fit within the lowest eight states in the L-shaped eight sites. This turns out to be seven states in the first iteration, so together with the free site we would have 14 states per nine-site block.  This simple scheme yields a remarkable value, $-0.66927$ as energy density, which is $+0.03\%$ from the exact value. With this encouraging result, we are ready to see what happens in the more complicated $J_1-J_2$ model.

\section{The $J_1-J_2$ model}
\label{sec:j1j2}

As we have seen, CORE's way of exhibiting the long-range N\'{e}el order of the HAF, is that RG with bootstrap always generates strong diagonal couplings; i.e., in all the blocking schemes $c_{xuxu}/(c_{xx}+2c_{xxuu})$ flows to a large negative number.  This raises the question of what happens if we add a large positive $c_{xuxu}$ at the initial step.  This, of course, brings us to the well known frustrated $J_1-J_2$ model:
\be
H= J_1 \sum_{nn} \vec{S}\cdot \vec{S} +
J_2 \sum_{diag} \vec{S}\cdot \vec{S}.
\ee
where the first sum is over nearest-neighbors and the second sum is over diagonal sites.  This corresponds to $\{c_{xx}=J_1, c_{xuxu}=J_2/J_1, c_{xxuu}=c_{xzxz}=c_{xxzz}=0\}$.  The $J_2-J_1$ model has been extensively studied in the condensed-matter literature (\cite{sow,j1j2,j1j2ent} and references therein) and it is generally believed that there are several phase transitions within the region $0<J_2/J_1<1$.  Ref.\cite{sow} for example lists four conjectured critical points at $J_2/J_1=0.34, 0.38, 0.5, 0.62$.  In this work  the ground state transitions from the N\'{e}el phase to various dimerized configurations with increasing frustration .

Past works on this subject typically rely on two types of techniques. One is to use perturbative methods around some conjectured starting configurations; the other is to perform calculations on finite lattices with some numerical methods and extrapolate to an infinite lattice. The bootstrap CORE is therefore a very different method and may provide independent checks on past results.  Since we have an RG flow picture that describes the N\'{e}el phase at $J_2/J_1=0$, we would expect to find at least one critical point where this picture begins to change.

\begin{figure}
  \begin{center}
  \includegraphics[width=3.4in]{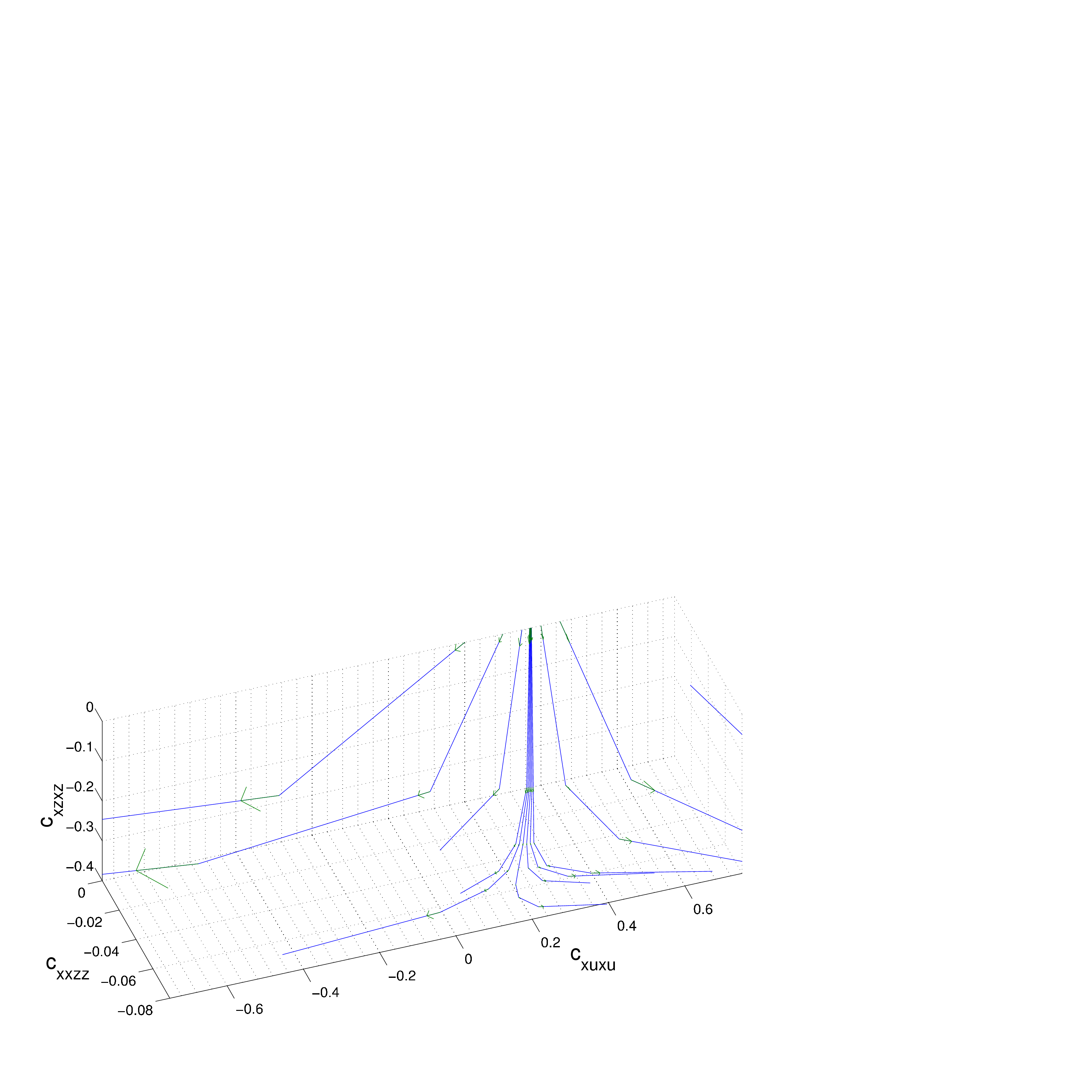}
  \end{center}
  \caption{(Color Online) Renormalization group flow generated by "8+1" bootstrap CORE for weak frustration $J_2/J_1 = \{0.2, 0.3, 0.35, 0.371, 0.372,
0.373, 0.374, 0.375, 0.39, 0.4, 0.45\}$. All the couplings above are
  implicitly divided by $(c_{xx}+2c_{xxuu})$, so that the overall scale does not affect our analysis of phases.
  We run the system starting from several values of $c_{xuxu}$ from 0.2 to 0.45; each arrow on the lines indicate a CORE iteration,
  their size indicates amount of change in each step. To avoid cluttering, this region in parameter space has been selected to give
  a clear picture of the quantum phase transition. }
  \label{frusrg1}
\end{figure}

In what follows, using the "8+1" bootstrap scheme, we compute the
renormalization flow for $J_2/J_1<0.5$, as shown in
Fig.\ref{frusrg1}. For $J_2/J_1\leq 0.371$, the system flows towards
the unfrustrated case, whereas $J_2/J_1 \geq 0.372$, the frustration
grows.  The qualitative feature of the entire region can be seen
quite clearly from the plot.  To our knowledge this is the first
time such a renormalization group result has been obtained directly
from first principles for the $J_2-J_1$ model.  In Ref.\cite{sow},
$J_2/J_1=0.34$ is thought to be the point where columnar spin dimers
begin to appear. Another critical point at $J_2/J_1=0.38$, which is
more established, is thought to be where the system becomes a
disordered spin-liquid. In between these two values is a region
where columnar dimer order and N\'{e}el order coexist.  The critical
point we see in Fig.\ref{frusrg1} appears to be in agreement with
literature.

\begin{figure}
  \begin{center}
  \includegraphics[width=3.4in]{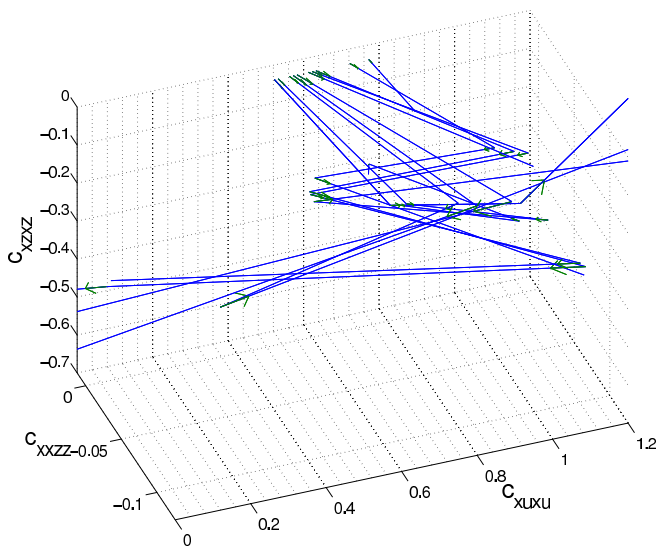}
  \end{center}
  \caption{(Color Online) Renormalization group flow starting with strong frustrations. The gap between the lowest spin-$\frac{1}{2}$
  and other states in the nine-site block is often small or even negative in much of this region, so our algorithm
  may not be reliable here. No clear pattern emerges compared to Fig.\ref{frusrg1}, and there are many large erratic
  jumps from strong to weak frustration region.}
  \label{frusrg2}
\end{figure}

Things do not go as well for $J_2/J_1\geq 0.5$.  As the frustration
increases, there is more competition among states and the gaps
decrease and the gap between the lowest spin-$\frac{1}{2}$ and other
excited states in the nine-site block become very small.  In fact,
after one iteration with $J_2/J_1=0.59$ or $J_2/J_1=0.75$, we arrive
at a system where the nine-site block no longer has a
spin-$\frac{1}{2}$ as its ground state, and so the truncation scheme
we are using should change.  If we don't do this then we see that,
starting with $J_2/J_1\geq 0.372$, the frustration eventually grows
to a point where the CORE recipe breaks down.  Fig.\ref{frusrg2}
shows this happening to the flow of couplings at $J_2/J_1\geq 0.5$.

\begin{figure}
  \begin{center}
  \includegraphics[width=3.4in]{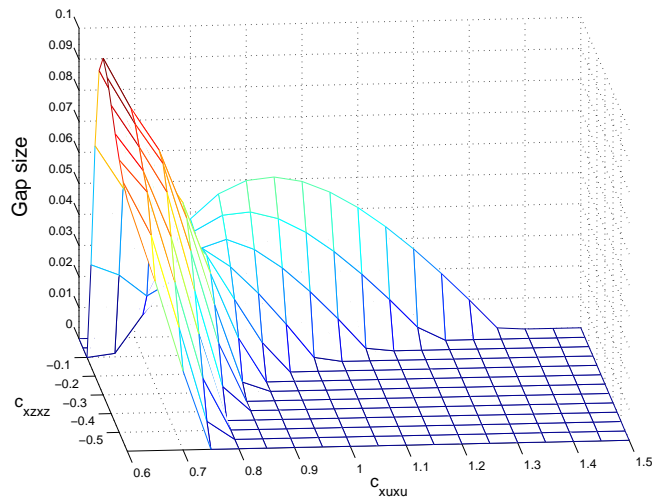}
  \end{center}
  \caption{(Color Online) Plot of the gap between the lowest spin-$\frac{1}{2}$ multiplet and the first excited states in the
  nine-site block in a typical frustrated region. The couplings are implicitly in units of
  $(c_{xx}+2c_{xxuu})$, with $c_{xx}=1, c_{xxuu}=-0.2, c_{xxzz}=0$. The gap is set to zero when the spin-$\frac{1}{2}$
  multiplet ceases to be the ground states.}
  \label{gapplot}
\end{figure}

While the first iteration always takes us to higher frustration for
$0.372<J_2/J_1<\sim 0.9$, the flow is quite erratic in the strong
frustration region.  This presumably occurs because we fail to
modify our truncation procedure as needed.  Another curious behavior
that is probably also due to our failure to follow the true ground
state, is that there is a neighborhood around $c_{xuxu} \sim 1$
where the flow takes us back to weakly frustrated regions.  For
example, if we start with $\{c_{xuxu},c_{xxzz}, c_{xzxz}\}=\{1.085,
-0.064, -0.223\}$, we would be taken to $\{0.318,-0.021,-0.086\}$ at
the next iteration, and small changes at the beginning lead to very
different couplings later on.  To give an idea how serious the gap
issue is, we plot in Fig.\ref{gapplot} the size of the first excited
state gap in the nine-site block as a function of
$c_{xuxu}/(c_{xx}+2c_{xxuu}), c_{xzxz}/(c_{xx}+2c_{xxuu})$ with
$c_{xx}=1, c_{xxuu}=-0.2, c_{xxzz}=0$.  All of these results are
consistent with what has been seen in one-dimensional examples when
one fails to modify the space of retained states as needed whenever
level crossings occur.

\begin{figure}
  \begin{center}
  \includegraphics[width=3.3in]{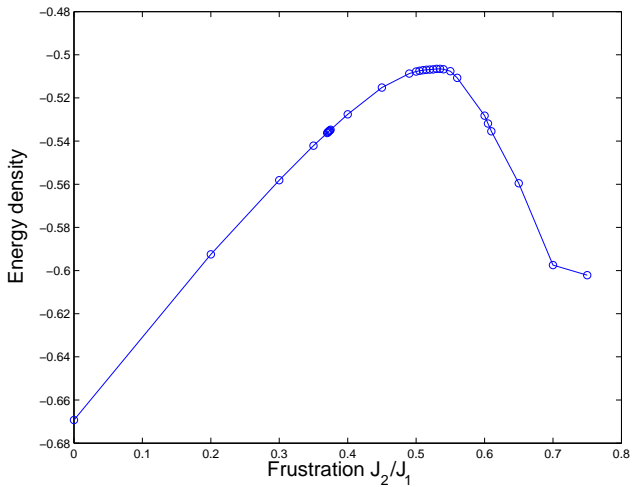}
  \end{center}
  \caption{Energy density as a function of frustration in the $J_1-J_2$ model. The slope appears to be the steepest around $0.6$.
  The last data point at $0.75$ should be lower than it appears since we have to stop after one iteration.}
  \label{frusenergy}
\end{figure}

Compared to the couplings, we see that, because it is largely
determined by the first two iterations, the energy density is a
little less sensitive to the small gaps.  An plot of the energy
density with respect to $J_2/J_1$ is shown in Fig.\ref{frusenergy}.
This can be contrasted with Fig.5 of Ref.\cite{dmrgmc}, where a
qualitatively similar plot is obtained from finite lattice
extrapolation. Instead of a peak and discontinuity at
$J_2/J_1\sim0.6$, we find a peak around $0.53$. No particular
structure is found near $J_2/J_1\sim0.6$, but the slope appears to
be the steepest there.  It is possible that these features are
indirect signs of conjectured phase transitions at $J_2/J_1=0.5,
0.6$ (similar features are found in finite lattice entropy plots
\cite{j1j2ent}), but we would have to run magnetization and other
specific operators such as columnar dimers \cite{sow} to
characterize the phases in detail.

\section{Performance of Other Variants}
\label{sec:other}

In conclusion, we would say that our bootstrap algorithm appears to
be a promising way of increasing the accuracy of CORE computations.
However, we don't have a general prescription for estimating the
number of states to keep in the buffering step.  Moreover, we
haven't yet tried to do the more difficult questions of how to
handle the truncation scheme when we flow to points where level
crossings occur.

The issue of strong frustration aside, there are other questions
which need more study.  How robust are our results? What specific
features in the blocking scheme work best?  How do we characterize
the different phases believed to exist in the $J_1-J_2$ model, if
the flow of the coefficients in the Hamiltonian fail to clearly
distinguish them?  In this section we will discuss some preliminary
results on variants of the algorithm and the puzzles they pose.

\subsection{Use of Reduced Density Matrix}

Under the $8+1$ bootstrap scheme the retained states on the eight
sites are chosen according to their energies. One might question the
arbitrariness of favoring the energy spectrum of this irregular
eight-site configuration. One alternative is to adopt the reduced
density matrix truncation technique from the DMRG
\cite{densitymatrix} literature. The use of reduced density matrix
in CORE first appeared in Ref.\cite{capponi1}, where it was
suggested as a diagnostic tool. Ref.\cite{haf1} explored this idea
further and argued that while this diagnostic tool is not
necessarily reliable (since CORE does not rely solely on
preservation of state structure - see Ref.\cite{haf1} for comparison
between DMRG and CORE), it can provide an alternative truncation
method. This method has yielded interesting results in Refs.
\cite{haf1,capponi2}, so it is instructive to try it on the $8+1$
bootstrap calculation.

The truncation method, in the present context, means the following.
We first consider a site configuration (a \emph{superblock} if we
follow DMRG terminologies) that contains the eight sites of
interest, i.e. a configuration that is the eight sites plus some
"environment" sites. On the superblock, we will compute a target
state, which can be the ground state of the superblock or a mixed
state formed by some low energy states. Denoting the target state
density matrix as $\sigma$ and the environment as $E$, we compute
the reduced density matrix on the eight sites as follows: \be \rho =
\Tr_E \sigma. \ee (If the target state is a pure state $\p$,
$\sigma=\p\bra{\psi}$.) The eigenvectors corresponding to the
largest eigenvalues of this matrix would tell us what states in the
eight sites contribute the most to the formation of $\p$, and so
this gives us a way to choose what states to keep.

It should be apparent that the result we get would strongly depend
on how we choose the target state (and the environment). To make
things as simple as possible in our test, we choose the superblock
to be a 16-site square and the target state to be its spin-0 ground
state. This way we don't have to worry about degeneracies or
enforcement of spin symmetry. Approximately seven eigenstates of
$\rho$ (depending on the multiplet structure) with the largest
eigenvalues are chosen as the retained states. With this particular
choice of target state, the $8+1$ bootstrap CORE for the
unfrustrated HAF gives us an energy density of -0.66962, which is
$-0.03\%$ from the exact value and as good as our original recipe.

Unfortunately, this choice does not seem to perform as well in
identifying phase transitions in the $J_1-J_2$ model. In
Ref.\cite{haf1} we have seen that targeting the superblock ground
state could lead to improvement in ground state energy at the
expense of accuracy in excited states. Thus we might expect the
target state choice to be related to a tradeoff between the ground
state energy and the overall physical picture. Indeed, what we
observe here is a subtly altered RG flow. For relatively weak
frustration, say, $J_2/J_1=0.3$, we see an initial flow to the left
of Fig.\ref{frusrg1}, i.e. the less frustrated region.  Yet as we
enter the far front left region (negative $c_{xuxu}$ and negative
$c_{xxzz}$), the renormalization turns around and changes direction
toward strong frustration on the right.

It is possible, of course, that the situation would improve if we
use a mixed target state that targets some states above the ground
state of the superblock. This technique could be important for not
just the choice of states for the bootstrap step, but also for the
choice of final retained states in the nine-site block.  We have
seen in Fig.\ref{gapplot} that the lowest spin-$\frac{1}{2}$ may not
sufficiently capture the low energy spectrum under strong
frustration. If we would like to continue to work with
spin-$\frac{1}{2}$, the reduced density matrix approach may be the
only option.

\subsection{Resummation}
\label{sec:resum}

There is another degree of freedom in the procedure that affects the choice of retained states, and that is how we decide which part of the Hamiltonian belongs to a given set of sites.  Suppose we want to include all the operators that act on two bodies A and B, and we know there is an operator $O_4$ that acts on four bodies A, B, C and D.  For any operator $O_2$ that acts on A and B, we can always decompose $O_4$ into $(O_4-O_2)+O_2$.  Hence it should be equally valid to say that we have an operator $O_4-O_2$ acting on the four bodies, and an operator $O_2$ that acts on A and B. But $O_2$ can be anything, so what should we choose it to be?

In our case, the question is whether we should set $c_{xx} \leftarrow c_{xx}+2c_{xxuu}$, $c_{xxuu} \leftarrow 0$, i.e. consider the $c_{xxuu}$ term as a truly two body term instead of part of a four body term.  This would seem like a very natural thing to do, but it would dramatically change the cluster expansion.  Running $8+1$ bootstrap CORE with this resummation yields $-0.66957$ for the HAF energy density, and the growing ferromagnetic diagonal couplings come out as well.  However, the behavior of couplings in the $J_1-J_2$ model turns out to be very different.  This time, instead of running toward strong frustration too early, the system runs to N\'{e}el for relatively large $J_2/J_1$.  The critical point shifts to above $0.4$.  This change may not seem surprising if we note that applying resummation to the five-site blocks in Ref.\cite{haf1} also obliterates the growing spin picture there. The problem is how we should interpret the effects of resummation.

Mathematically the issue of grouping terms in the Hamiltonian is hardly different from the issue of selecting a blocking scheme, which strongly affects the resulting qualitative picture in CORE. One possible explanation for the effect of resummation is that the original grouping contains some structural information of the model in large scale, such as what counts as self-interactions within a group of blocks and what counts as long-range interactions. Since we would like to study the flow of parameters in the same parameter space, it may be more consistent to preserve this information.  Under this interpretation, resummation can yield very accurate energies, but we may lose control on the qualitative picture as a result.

\section{Further Discussion}
\label{sec:disc}

We have shown that it is possible to approximate long-range terms in
CORE using a bootstrap approach and a relatively small number of additional states.  In particular, we have obtained some remarkable renormalization flows that illuminate the phases of a 2-D antiferromagnet.  However, an equally important lesson that can be drawn from our results is that CORE is sensitive to blocking and truncation schemes.  This suggests that the method itself needs further study.  Moreover, it is clear that one should be careful about drawing any strong conclusions about specific applications without checking that the important results one arrives at are common to several different schemes.

For future directions, our work should be immediately extended in two ways.  The first is to investigate the effect of using reduced density matrix truncation with different target states at both stages of the bootstrap approximation. This has the prospect of resolving symmetry problems and allowing us to smoothly transition between truncation schemes when level crossings occur.  This would be particularly important in extending our results to the highly frustrated case. Second, of course, is that we have not attempted to study how the expectation values of various relevant operators change in the $J_2-J_1$ model with increasing frustration.  Such information is crucial to shedding more light on the structure of the phases.


\end{document}